\documentstyle[aps,12pt]{revtex}
\newcommand{\bb}{\begin{equation}}
\newcommand{\ee}{\end{equation}}
\newcommand{\ba}{\begin{array}}
\newcommand{\ea}{\end{array}}

\begin {document}
\baselineskip 2.2pc

\title{Levinson theorem for Dirac particles in one dimension\thanks
{published in Eur. Phys. J. D {\bf 7} (1999) 515-524.}}
\author{Qiong-gui Lin\thanks{E-mail addresses: 
        qg\_lin@163.net, stdp@zsu.edu.cn}}
\address{China Center of Advanced Science and Technology (World
	Laboratory),\\
        P.O.Box 8730, Beijing 100080, People's Republic of China\\
        and\\
        Department of Physics, Zhongshan University, Guangzhou
        510275,\\
        People's  Republic of China \thanks{Mailing address}}

\maketitle
\vfill

\begin{abstract}
{\normalsize
The scattering of Dirac particles by symmetric potentials in one
dimension is studied.
A  Levinson theorem is established. By this theorem, the 
number of bound states with even (odd) parity, $n_+$ ($n_-$),
is related to the phase shifts  $\eta_+(\pm E_k)$ [$\eta_-(\pm E_k)$]
of scattering states with the same parity at zero momentum as follows:
$$\eta_\pm(\mu)+\eta_\pm(-\mu)\pm{\pi\over 2}[\sin^2\eta_\pm(\mu)
-\sin^2\eta_\pm(-\mu)]=n_\pm\pi.$$
The theorem is verified by several simple examples.}
\end{abstract}
\vfill

\leftline {PACS number(s): 34.10.+x, 03.65.-w, 11.80.-m}
\newpage

\baselineskip 15pt

\section{Introduction}               %I

In 1949, Levinson established a theorem in nonrelativistic quantum
mechanics [1]. The theorem gives a
relation between bound states and scattering states in a given
angular momentum channel $l$, i.e.,  the total number of bound states
$n_l$ is related to the phase shift $\delta_l(k)$ at threshold
($k=0$):
$$\delta_l(0)=n_l\pi,\quad l=0,1,2,\ldots.\eqno(1{\rm a})$$
The case $l=0$ should be modified as
$$\delta_0(0)=(n_0+1/2)\pi\eqno(1{\rm b})$$
\addtocounter{equation}{1}
when there exists a zero-energy resonance (a half-bound state) [2].
This is one of the most interesting
and beautiful results in nonrelativistic quantum theory. The subject
was then studied by many authors (some are listed in the Refs. [2-8])
and  generalized to relativistic quantum mechanics [6,9-15].
In Ref. [6] the theorem was first written as
\bb
\delta_l(0)-(-)^l{\pi\over2}\sin^2\delta_l(0)=n_l\pi.
\ee       %2
The second term on the left-hand side (lhs) appears automatically
in the Green function approach to Levinson's theorem, which was
used in earlier works [3,4] and further developed by Ni [6]. This
method is quite different from the original one where the theorem
is obtained by using the analytic properties of the Jost functions,
and is convenient for generalization to the relativistic case.
It can be shown that $\delta_l(0)/\pi$ takes integers when $l>0$
or when $l=0$ but there exists no zero-energy solution. Then Eq. (2)
is equivalent to Eq. (1a). When $l=0$ and there exists a zero-energy
solution (a half-bound state), however, $\delta_l(0)/\pi$ takes
half integers and $\sin^2\delta_0(0)=1$. In this case Eq. (2) is
equivalent to Eq. (1b). Thus Eq. (2) combines the two cases in
Eq. (1) into a unified form. The correct
generalization of Levinson's theorem to Dirac particles was first
obtained by the Green function method in Ref. [10]. If the phase
shifts in the angular momentum channel $\kappa$ ($\kappa=\pm1,
\pm2,\ldots$) are denoted by $\eta_\kappa(\pm E_k)$ and the total
number of bound states in that channel by $n_\kappa$, then the
Levinson theorem read
\bb
\eta_\kappa(\mu)+\eta_\kappa(-\mu)-(-)^\kappa\epsilon(\kappa)
{\pi\over 2}[\sin^2\eta_\kappa(\mu)
-\sin^2\eta_\kappa(-\mu)]=n_\kappa\pi,
\ee              %3
where $\epsilon(\kappa)=1$ ($-1$) for positive (negative) $\kappa$,
$\mu$ is the mass of the particle. The effect of half-bound states
has been automatically included in this equation, though the problem
is more complicated than in the nonrelativistic case.
There is a modulo-$\pi$ ambiguity in the definition of the phase
shifts. This is common to both nonrelativistic and relativistic
cases. In the former case this was resolved by setting $\delta_l
(\infty)=0$ (rather than a multiple of $\pi$)
which can be freely done.
In the latter case one is not allowed to set $\eta_\kappa(\pm\infty)
=0$, but the modulo-$\pi$ ambiguity can be appropriately
resolved [9, 10, 16 17].

Most of the above cited works mainly deal with
the problem in ordinary three-dimensional space. 
Recently, several authors have studied the two-dimensional version
of Levinson's
theorem for both nonrelativistic Schr\"odinger particles and
relativistic Dirac particles [18-20]. In the works of the present
author the Green function method is employed [18].
In the
nonrelativistic case we obtain 
\bb
\delta_m(0)=n_m\pi, \quad m=0,1,2,\ldots.
\ee                %4
In two dimensions 
$\delta_m(0)/\pi$ always take integers. This threshold behavior is
quite different from that in three dimensions
and in one dimension (see below). In Ref. [19], Dong, Hou, and Ma
studied the problem by the method of Sturm-Liouville theorem and
found that the case $m=1$ in Eq. (4) should be modified as
\bb
\delta_1(0)=(n_1+1)\pi
\ee                %5
when there exists a half-bound state in this channel. In other
words, the half-bound state with $m=1$ plays the role of a real
bound state in the Levinson theorem. The same result was also
obtained  in an earlier paper [20]. Similar modifications also occur
in the relativistic case. The miscounting in our work may be due to
the very subtle behavior of the above state and the shortcomings of
the regularization procedure we employed [cf. Eqs. (44-47) in this
paper], rather than the failure of the completeness relation as
remarked in Ref. [19]. (see some more discussions in Sec. II.) The
above state is quite different from the half-bound state with $m=0$
and from the half-bound states in three and one dimension because it
tends to zero at infinity. As
a result it is lost in the regularization procedure. In three- and
one-dimensional space there is no similar case and the Green function
method leads to correct results.

It seems that the one-dimensional version of Levinson's theorem has
attracted more attention than the two-dimensional one. In fact,
the nonrelativistic case has been studied by several authors [21-23].
In a symmetric potential $V(x)$ (an even function of $x$), 
the Levinson theorem takes the form
\bb
\delta_\pm(0)\pm{\pi\over2}\sin^2\delta_\pm(0)=n_\pm\pi,
\ee        %6
where + ($-$) indicates even (odd) parity.
It is easily seen that the odd-parity case coincides with the case
$l=0$ in three dimensions, while the even-parity case 
has no counterpart in three dimensions. This is the main feature
in one dimension.

The purpose of this paper is to generalize the one-dimensional
Levinson theorem to relativistic Dirac particles.
The interest of this problem is threefold. First,
the Levinson theorem is a nonperturbative result. It may be useful
in the study of nonperturbative field theories.
In view of the wide interest in field theories
and condensed matter physics in
lower dimensions in recent years, and the potential applications
of the Levinson theorem, the problem seems of interest. Indeed,
some applications
of the theorem to field theories have appeared in the literature
[21, 24, 25]. More recently,
the Levinson theorem in two dimensions has been used
[26] to the study of the screened Coulomb potential which
plays an important role in the physics of semiconductor
heterostructures.
Second, as an academic problem, it exhibits some new features
as compared with the three- or two-dimensional problem. 
Thus it may be of some interest in itself, as many one-dimensional
models, say, the one-dimensional QED (the Schwinger model), studied
in the literature. Third, it may be related to the real
three-dimensional problem. Consider a Dirac particle in an external
potential $V$ (the zero component of a vector potential). If $V$
depends only on one space variable, say, $x$, one may consider a
special case where the particle moves only in the $x$ direction.
This special case may be described by the one-dimensional Dirac
equation. This connects the problem to the real physical world.

Consider Dirac
particles in a symmetric potential in one dimension. 
The phase shifts of even-(odd-)parity solutions (see Sec. II)
are denoted by $\eta_+(\pm E_k)$ [$\eta_-(\pm E_k)$], while the
number of bound states with the same parity is denoted by $n_+$
($n_-$). The Levinson theorem connects them with each other:
\bb
\eta_\pm(\mu)+\eta_\pm(-\mu)\pm{\pi\over 2}[\sin^2\eta_\pm(\mu)
-\sin^2\eta_\pm(-\mu)]=n_\pm\pi.
\ee     %7
We will establish this theorem for cutoff potentials (vanish when
$|x|>a>0$) in this paper. Though the even-(odd-)parity
case coincides in appearance with the case $\kappa=1$ ($\kappa=-1$)
in three dimensions [cf. Eq. (3)]. They are essentially different.
First, the one-dimensional equation
is different from the radial one with $\kappa=1$ or $\kappa=-1$ in
three dimensions. The boundary conditions at the origin are also
different. Second, the two critical energy solutions with
$E=\pm\mu$ and even (odd) parity, if exist, are both half-bound
states in one dimension, while one of the two in three dimensions
with $\kappa=1$ ($\kappa=-1$) is a bound state. Third, the threshold
behaviors of the phase shifts are different. For example, in three
dimensions $\eta_1(\mu)/\pi$ always takes integers, while in one
dimension $\eta_+(\mu)/\pi$ takes integers only when there exists a
half-bound state with $E=\mu$ and even parity. In general $\eta_+
(\mu)/\pi$ takes half integers. There is a similar difference between
$\eta_-(-\mu)$ and $\eta_{-1}(-\mu)$. Therefore each case in Eq. (7)
has no counterpart in three dimensions.

Throughout this paper we use natural units where $\hbar=c=1$.
In the next section we first discuss the solutions
of the one-dimensional Dirac equation in an external symmetric
potential. Then we discuss the scattering problem. 
In Sec. III the behavior of the phase shifts near $k=0$ is analyzed.
The behavior when there exists a half-bound state is distinguished
from the case without one. 
In Sec. IV we establish the Levinson theorem for cutoff potentials.
In this paper we do not
resort to the Green function method developed in Ref. [6] and used in
Refs. [10, 17-18].  We employ the completeness of the solutions of the
Dirac equation  to derive the
theorem more directly. This is essentially equivalent to the Green
function method but simpler, and has been used in some recent works
[13, 23]. In Sec. V, we 
first discuss the behavior of the
phase shifts at infinite momentum and the resolution of the
modulo-$\pi$ ambiguity in the definition of the phase shifts. Then
we examine the Levinson theorem by two simple examples. 
Finally we give a brief summary of the results.

\section{Dirac particles in one dimension}          %II   

We work in (1+1)-dimensional space-time. The Dirac equation in an
external vector field $A_\nu(t,x)$ reads
\begin{equation}
(i\gamma^\nu D_\nu-\mu)\Psi=0,
\end{equation}                  %8
where $\mu$ is the mass of the particle, $D_\nu=\partial_\nu+ieA_\nu$,
$e$ is the coupling constant, and summation over the repeated Greek
index $\nu$ ($\nu=0,1$) is implied. The $\gamma^\mu$ are Dirac
matrices satisfying the Clifford algebra:
\begin{equation}
\{ \gamma^\mu, \gamma^\nu \}=2g^{\mu\nu},
\end{equation}       %9
where $g^{\mu\nu}={\rm diag}(1,-1)$ is the Minkowskian metric.
In this paper we only consider the zero component of $A_\nu$,
which is an even function of $x$, namely, we consider the special
case where
\begin{equation}
A_1=0, \quad eA_0=V(x),
\end{equation}              %10
where $V(x)$ is symmetric with respect to reflection:
\bb
V(-x)=V(x).
\ee        %11
In this case we may set
\begin{equation}
\Psi(t,x)=e^{-iEt}\psi(x),
\end{equation}          %12
and get a stationary equation for $\psi(x)$:
\begin{equation}
H\psi=E\psi,
\end{equation}              %13
where the Hamiltonian
\begin{equation}
H=\alpha p+\gamma^0\mu+V(x),
\end{equation}           %14
where $p=-i\partial_x$ is the momentun operator and $\alpha=\gamma^0
\gamma^1$.

To solve the Dirac equation (13) an explicit representation of the
Dirac matrices is necessary. In one dimension
this can be realized by the Pauli matrices:
\begin{equation}
\gamma^0=\sigma^3,\quad \gamma^1=i\sigma^1.
\end{equation}                     %15
In this representation $\alpha=-\sigma^2$. We denote the two-component
spinor $\psi(x)$ as
\begin{equation}
\psi(x)=\left(
\begin{array}{c}
u(x)\\    v(x) \end{array}
\right)=\left(
\begin{array}{c}
u_1(x)\\  u_2(x) \end{array}
\right),
\end{equation}      %16                        
where the second notation will be used only in a few occasions where
summation over the spinor index is involved, then Eq. (13) can be
explicitly written as a system of first-order differential equations
for $u$ and $v$:
\bb
u'+(E+\mu-V)v=0,\quad
v'-(E-\mu-V)u=0.
\ee     %17

Since $V(x)$ is an even function of $x$, it is easy to show that if
$(u(x),v(x))^\tau$ is a solution of Eq. (17), then
$(u(-x),-v(-x))^\tau$ is also a solution with the same energy $E$,
where $\tau$ denotes matrix transposition. Thus the solutions of
Eq. (17) can be chosen to have definite parities. Even-parity
solutions are denoted by $\psi(x,+)$ and defined by the property
of its components under reflection:
$$ u(-x, +)=u(x,+),\quad v(-x,+)=-v(x,+).\eqno(18{\rm a})$$
Odd-parity solutions are denoted by $\psi(x,-)$ and defined by
$$ u(-x, -)=-u(x,-),\quad v(-x,-)=v(x,-).\eqno(18{\rm b})$$

If $(u(x),v(x))^\tau$ and $(\tilde u(x),\tilde v(x))^\tau$ are both
solutions to Eq. (17) with the same energy value $E$, it can be
shown that
$u\tilde v-\tilde u v={\rm constant}$.
For bound states, all the functions $u$, $v$, $\tilde u$, and
$\tilde v$ must vanish at infinity, so the constant in the above
equation is zero, and we have $\tilde u/u=\tilde v/v$. We denote this
fraction by $w$, a function of $x$, and get $\tilde u=wu$, $\tilde v
=wv$. Substituting this into Eq. (17), we have $w'u=0$, $w'v=0$. As
a nontrivial solution, $u$ and $v$ cannot simultaneously vanish at any
point, so we have $w'=0$, or $w={\rm constant}$,
which means that the two
solutions are equivalent. Thus there is no degeneracy with bound
states.

From the above discussions we conclude that a bound state solution of
Eq. (17) must have definite parity. On the other hand, scattering
states need not vanish at infinity. For a given energy, there are two
linearly independent solutions. Both are physically acceptable. They
do not necessarily have definite parities. However, they can be chosen
such that one is of even parity and the other of odd. The above
discussion is applicable to the case of free particles where $V=0$
and there is no bound state.

For free particles, $V=0$, then Eq. (17) can be easily solved.
We have positive-energy solutions with $E>\mu$ and
negative-energy solutions with $E<-\mu$,
both being scattering solutions. We define
$k=\sqrt{E^2-\mu^2}\ge0$, and denote positive-(negative-)energy
solutions by the subscript $k$ ($-k$), thus we have, say,
\addtocounter{equation}{1}
\begin{equation}
E_{\pm k}=\pm E_k=\pm \sqrt{k^2+\mu^2}.
\end{equation}           %19
The solutions will be given below.
It is remarkable that when $E=\mu$ there is a  nontrivial
solution (not identically zero) with  even-parity
while when $E=-\mu$ there is one with odd-parity.
These are uaually called half-bound states.

Now we consider particles moving in the external symmetric potential
$V(x)$. In this section the potential need not be a cutoff one.
We assume that $V(x)$ decreases more rapidly than $x^{-2}$
when
$x\to \infty$, is less singular than $x^{-1}$ when $x\to 0$,
and is regular everywhere (except possibly at
$x=0$). 
It can be shown that $E^2>\mu^2$ gives scattering solutions
while $E^2<\mu^2$ gives bound state solutions. (The critical energy
states with $E=\pm \mu$ will be discussed in the next
section.) Scattering states will be
denoted as above. Bound states with even (odd) parity will be denoted
by a subscript $\kappa_+$ ($\kappa_-$) which takes discrete values.
Scattering states are normalized by Dirac $\delta$ functions.
These are not necessary for the following discussions
and are thus omitted. The orthonormal relation for bound states is
given by
\bb
(\psi_{\kappa'_\pm},\psi_{\kappa_\pm})
=\int_{-\infty}^{+\infty} dx\,
\psi_{\kappa'_\pm}^\dagger(x)\psi_{\kappa_\pm}(x)
=\delta_{\kappa'_\pm \kappa_\pm},
\ee      %20
which will be used below. Since the Hamiltonian (14) is an Hermitian
operator, we have the completeness relation
\begin{eqnarray}
&&\int_0^\infty dk\,[\psi_k(x,+)\psi_k^\dagger(x',+)+
\psi_k(x,-)\psi_k^\dagger(x',-) \nonumber \\
&&+\psi_{-k}(x,+)\psi_{-k}^\dagger(x',+)+
\psi_{-k}(x,-)\psi_{-k}^\dagger(x',-)] \nonumber\\
&&+\sum_{\kappa_+}\psi_{\kappa_+}(x)\psi_{\kappa_+}^\dagger(x')
+\sum_{\kappa_-}\psi_{\kappa_-}(x)\psi_{\kappa_-}^\dagger(x')
=\delta (x-x').
\end{eqnarray}          %21
It should be remarked that all regular solutions should be included
in the summation or integration. Thus half-bound states, when they
appear, should be included, or the above equation does not hold.
This can be verified straightforwardly
in the free case where two half-bound states are present as pointed
out above.
The asymptotic forms of the various types of scattering
sulutions are given by
$$
u_{\pm k}(x,+)\to\pm\sqrt{E_k\pm \mu\over 2\pi E_k}
\cos [k|x|+\eta_+(\pm E_k)],\eqno(22{\rm a})$$
$$
v_{\pm k}(x,+)\to \sqrt{E_k\mp \mu \over 2\pi E_k}\epsilon(x)
\sin [k|x|+\eta_+(\pm E_k)], \eqno(22{\rm b})$$
$$
u_{\pm k}(x,-)\to \sqrt{E_k\pm \mu \over 2\pi E_k}\epsilon(x)
\sin [k|x|+\eta_-(\pm E_k)], \eqno(22{\rm c})$$
$$
v_{\pm k}(x,-)\to \mp \sqrt{E_k\mp \mu\over 2\pi E_k}
\cos [k|x|+\eta_-(\pm E_k)],\eqno(22{\rm d})$$
\addtocounter{equation}{1}
when $x\to\infty$ ($+\infty$ or $-\infty$), where $\epsilon(x)=1$
($-1$) for positive (negative) $x$, $\eta_+(\pm E_k)$
[$\eta_-(\pm E_k)$] are the phase shifts of even-(odd-)parity
solutions. The phase shifts depend on the sign as well as the
magnitude of the energy,
as Eq. (17) does.
By setting all the phase shifts to zero, the right-hand side
(rhs) of Eq. (22) gives exactly the solutions for free particles.
Compared with the free solutions, the asymptotic
forms in the external symmetric potential are distorted by the phase
shifts. But it should be remarked that the normalization factors
in Eq. (22) are the same as in the free case.

It is well known that positive-(negative-)energy solutions
correspond to particles (antiparticles) after second quantization.
In the following we give some results for the scattering of
positive-energy
solutions by the symmetric potential $V(x)$ described above. The
results for the
scattering of negative-energy solutions are similar.

We denote the reflection amplitude by $R_k$ and the transmission one
by $T_k$. As a consequence of charge conservation,
they satisfy
\bb
|R_k|^2+|T_k|^2=1.
\ee      %23
Similar to the partial-wave method in three or two dimensions,
$R_k$ and $T_k$ can be expressed in terms of the phase shifts.
The result reads
$$
R_k={1\over2}[e^{2i\eta_+(E_k)}-e^{2i\eta_-(E_k)}]=i
e^{i[\eta_+(E_k)+\eta_-(E_k)]}\sin [\eta_+(E_k)-\eta_-(E_k)],
\eqno(24{\rm a})$$
$$
T_k={1\over2}[e^{2i\eta_+(E_k)}+e^{2i\eta_-(E_k)}]=
e^{i[\eta_+(E_k)+\eta_-(E_k)]}\cos [\eta_+(E_k)-\eta_-(E_k)].
\eqno(24{\rm b})$$
These results obviously satisfy Eq. (23). They hold for both
left-incident particles and right-incident ones since  the
potential is symmetric.
If the potential is not symmetric,
the reflection amplitude would depend on the direction of
incidence.
We can construct a scattering
matrix and discuss various properties of it as in the nonrelativitic
case [22]. However, we will not go further in this respect.
In the following we will
confine our discussions to symmetric potentials.

From the above discussions, we see
that information about the scattering process
is contained in the phase shifts. The latter are determined by
solving the system of equations (17) with the boundary conditions
(22), and thus depend on the particular form of the potential.
On the other hand, the total number of bound states with definite
parities also depends on the particular form of the potential.
The purpose of the Levinson theorem is to establish a
relation between the phase shifts and the total
number of bound states with a given parity.

\section{Phase shifts near threshold}      %III

In this section we discuss the behavior of the phase shifts
$\eta_\pm(E_k)$  and $\eta_\pm(-E_k)$ near $k=0$. This is important
in the development of the Levinson theorem, and is also helpful
in understanding the result.

From now on we consider cutoff potentials
that satisfy $V(x)=0$ when $|x|>a>0$.
Such potentials will be denoted by $V_a(x)$ in the following.
Since we will deal with solutions with definite parities, we need only
consider the region $x\ge0$. The solutions in the region $x>a$ will
be indicated by a superscript $>$. Scattering solutions are given by
those on the rhs of Eq. (22) since $V_a(x)=0$ in
this region. Bound state sulutions will be discussed below.
In the region $x<a$, the solutions are not explicitly available.
First we consider the behavior of the solutions near $x=0$ and set
appropriate boundary conditions. As
assumed in Sec. II, $V_a(x)$ is less singular than $x^{-1}$ when $x\to0$.
If $V_a(x)$ is regular at $x=0$, the
appropriate boundary conditions for Eq. (17) are obviously
$$
u(0,+)=1,\quad v(0,+)=0,\eqno(25{\rm a})$$
$$
u(0,-)=0,\quad v(0,-)=1.\eqno(25{\rm b})$$
If $V(x)$ behaves like $A_0/x^{1-\rho/2}$ when $x\to 0^+$
where $A_0$ is a constant and $0<\rho<2$,  it can be shown that
the above boundary conditions remain valid.
These boundary conditions are applicable
to both scattering solutions and bound state solutions, with positive
or negative energy values. Let us consider scattering solutions of
Eq. (17) with positive energy $E_k$ that satisfy the boundary
conditions (25). Note that Eq. (17) depends on $k^2$ rather than
$k$, and the boundary conditions (25) are independent of $k$. Thus
these solutions should only depend on $k^2$ as well. In other words,
they are even functions of $k$. We will denote them by $f_\pm(x,k^2)$
and $g_\pm(x,k^2)$,
where the subscript $+$ ($-$) indicates even (odd) parity.
Thus scattering  solutions with positive energy $E_k$ are given
in the region $x<a$ by
\addtocounter{equation}{2}
\bb
u_k^<(x,\pm)=A_\pm (k)f_\pm(x,k^2),\quad
v_k^<(x,\pm)=A_\pm (k)g_\pm(x,k^2),
\ee       %26
where the superscript $<$ indicates the region $x<a$, $A_\pm(k)$
are constants which in general depend on $k$ such that the solutions
in the two regions can be appropriately connected.
Obviously, $u_k(x,\pm)$ and $v_k(x,\pm)$  should be continuous at
$x=a$, so that the probability density and the probability
current density are continuous at the point.  The phase shifts
$\eta_\pm(E_k)$ are determined by this condition. The results are
given by
$$
\tan\eta_+(E_k)={\beta_+(\xi)-\tan\xi 
\over 1+\beta_+(\xi) \tan\xi},
\quad
\tan\eta_-(E_k)={1+\beta_-(\xi) \tan\xi\over
\tan\xi -\beta_-(\xi)},
\eqno(27{\rm a})$$
where $\xi=ka$ and $\beta_\pm(\xi)$ are defined by
$$
\beta_\pm(\xi)=\sqrt{E_k+\mu\over E_k-\mu}
{g_\pm(a,k^2)\over f_\pm(a,k^2)}.
\eqno(27{\rm b})$$
The above results show that the behavior of
$\eta_\pm (E_k)$ is
determined by that of $\beta_\pm(\xi)$ and ultimately by that of
$f_\pm(a,k^2)$ and $g_\pm(a,k^2)$.
The general dependence of $f_\pm(x,k^2)$ and $g_\pm(x,k^2)$ on $k$
may be very complicated since Eq. (17) depends on $k$ in a rather
complicated way. Fortunately, only the property of
$f_\pm(x,k^2)$ and $g_\pm(x,k^2)$ near $k=0$ is necessary for
our purpose.

We consider the limit $E=E_k\to\mu$ ($k\to 0$) of Eq. (17).
In this limit it takes
the following form to the first order in $k^2$:
$$
u'(x)+P(x, k^2)v(x)=0,\quad v'(x)+Q(x, k^2)u(x)=0,
\eqno(28{\rm a})$$
where
$$
P(x,k^2)=2\mu+{k^2\over 2\mu}-V(x),\quad
Q(x,k^2)=-{k^2\over 2\mu}+V(x).
\eqno(28{\rm b})
$$
The equations depend only on $k^2$. Their solutions that satisfy
the boundary conditions (25) should  depend only on $k^2$ as well,
since the boundary conditions do not depend on $k$.
These solutions will be denoted by $\tilde f_\pm(x,k^2)$ and
$\tilde g_\pm(x,k^2)$. Note that both $P(x,k^2)$ and $Q(x,k^2)$
are integral functions of $k$.
Thus a theorem of Poincar\'e tells us
that $\tilde f_\pm(x,k^2)$ and $\tilde g_\pm(x,k^2)$, which satisfy
$k$ independent boundary conditions,
are also integral functions of $k$. On the
other hand, Eq. (17) coincides with Eq. (28) in the limit
$E=E_k\to\mu$ ($k\to 0$).
Therefore, $f_\pm(x,k^2)$ and $g_\pm(x,k^2)$ must coincide with
$\tilde f_\pm(x,k^2)$ and $\tilde g_\pm(x,k^2)$ respectively
in the limit $k\to0$,
since they satisfy the same boundary conditions. Hence
we conclude that $f_\pm(x,k^2)$ and $g_\pm(x,k^2)$
are analytic functions of $k$ in the neighbourhood of $k=0$.
The above analysis holds
regardless of whether the potential is cutoff or not, and the
functions $f_\pm(x,k^2)$ and $g_\pm(x,k^2)$ are analytic in $k$
near $k=0$ for any
fixed $x$ in the interval $[0,+\infty)$,  not only in $[0,a]$.

We have shown that $f_\pm(a,k^2)$ and $g_\pm(a,k^2)$ are even
functions of $k$ and analytic near $k=0$. 
Therefore, when $k\to0$, the leading term(s) for
$g_\pm(a,k^2)/f_\pm(a,k^2)$ must be given by one of the following
forms:
\addtocounter{equation}{2}
\bb
{g_\pm(a,k^2)\over f_\pm(a,k^2)}\to \alpha_1^\pm\xi^{2l_1^\pm},
\quad \alpha_2^\pm\xi^{-2l_2^\pm},
\quad \alpha_4^\pm+\alpha_3^\pm\xi^{2l_3^\pm},
\quad (\xi\to 0)
\ee                 %29
where $l_1^\pm$, $l_2^\pm$, and $l_3^\pm$ are natural numbers,
$\alpha_1^\pm$, $\alpha_2^\pm$, $\alpha_3^\pm$ and $\alpha_4^\pm$
are nonzero constants. Consequently, the leading term for
$\beta_\pm(\xi)$ is given by one of the following forms:
\bb
\beta_\pm(\xi)\to \alpha_\pm\xi^{2l_\pm-1},
\quad \tilde\alpha_\pm\xi^{-(2\tilde l_\pm-1)},
\quad (\xi\to 0),
\ee              %30
where $l_\pm$ and $\tilde l_\pm$ are natural numbers, $\alpha_\pm$
and $\tilde\alpha_\pm$ are nonzero constants. Substituting this
into Eq. (27a) we have 
$$
\tan\eta_+(E_k)\to b_+\xi^{2p_+-1}
\quad{\rm or}\quad
\tilde b_+\xi^{-(2\tilde p_+-1)},
\quad(\xi\to 0), \eqno(31{\rm a})
$$
$$
\tan\eta_-(E_k)\to b_-\xi^{-(2p_--1)}
\quad{\rm or}\quad
\tilde b_-\xi^{2\tilde p_--1},
\quad(\xi\to 0), \eqno(31{\rm b})
$$
\addtocounter{equation}{1}
where $p_\pm$ and $\tilde p_\pm$ are natural numbers, $b_\pm$ and 
$\tilde b_\pm$ are nonzero constants. Thus $\tan\eta_\pm(E_k)\to 0$
or $\infty$ in the limit $k \to 0$ or $\xi \to 0$, and
$\eta_\pm(\mu)/\pi$ take integers or half integers.

In order to study the threshold behaviour of $\eta_\pm(E_k)$ more
specifically, we consider bound state solutions of Eq. (17) with
positive energy
\bb
E=E_\lambda=\sqrt{\mu^2-\lambda^2},\quad 0\le\lambda\le\mu.
\ee                 %32
The solutions satisfying the boundary conditions (25) in the region
$x<a$ will be denoted by $F_\pm(x,\lambda^2)$ and
$G_\pm(x,\lambda^2)$. They are even
functions of $\lambda$ as implied by the notations, since the
equations are invariant under the change $\lambda \to -\lambda$
and the boundary conditions are independent of $\lambda$. Note
that when $-\lambda^2$ is replaced by $k^2$, $E_\lambda$ becomes
$E_k$ and the solutions become the above scattering ones since
the boundary conditions are the same. Thus we have the useful
relations
\bb
F_\pm(x,-k^2)=f_\pm(x,k^2),\quad
G_\pm(x,-k^2)=g_\pm(x,k^2).
\ee         %33
In the region $x>a$, the solutions are explicitly available since
$V_a(x)=0$. They can be continuously connected to the interior
solutions (for $x<a$) only for some specific values of $\lambda$,
which determine the discrete energy eigenvalues of bound states.
This is quite different from the case of scattering states, where
interior and exterior solutions can be continuously connected for any
$k$ if the phase shifts are chosen according to Eq. (27). On the
other hand, given an energy value $E_\lambda$, a bound state with
this energy eigenvalue exists only when the potential $V_a(x)$
has a specific form such that the solutions in the two regions
can be connected continuously. Here we are interested in the case
of the critical energy $E=\mu$ ($\lambda=0$). The exterior solutions
are then given by (up to a normalization factor)
\bb
u_\mu^>(x,\pm)=1,\quad v_\mu^>(x,\pm)=0.
\ee           %34
They can be continuously
connected to the interior ones only when $V_a(x)$ takes some
specific form such that
\bb
G_\pm(a,0)=0.
\ee       %35
On account of Eq. (33), we conclude that critical energy solutions
with $E=\mu$ exist if and only if
\bb
g_\pm(a,0)=0.
\ee       %36
From Eq. (34) we see that the critical energy solutions, if exist,
are not bound states since they are not normalizable. Nevertheless,
the argument below Eq. (18b) applies since $v_\mu^>=0$. Thus the
solution with $E=\mu$ is not degenerate. In other words, the two
solutions with $E=\mu$ and different parities cannot appear
simultaneously for a given potential. As in the free case
these solutions are called half-bound states.

Now we easily realize that the first limit in Eq. (29) corresponds
to the existence of critical energy states with $E=\mu$. It can
then be verified that this corresponds to the first case in Eqs. (30)
and (31). The other cases in these equations correspond to the case
without the above critical energy states. Thus we conclude that
$\eta_+(\mu)/\pi$ [$\eta_-(\mu)/\pi$] takes integers (half integers)
when these exists a critical state with $E=\mu$ and even (odd)
parity, otherwise it takes half integers (integers). It is easy to
check that $|T_0|=1$ when there exists a half-bound state with
$E=\mu$ (even or odd), otherwise $T_0=0$ [cf. Eq. (24)].

In the above we have analysed the threshold behaviour of
$\eta_\pm(E_k)$ in detail. The threshold behaviour of
$\eta_\pm(-E_k)$ can be discussed in a parallel way. 
The results are given by
$$
\tan\eta_+(-E_k)\to d_+\xi^{-(2q_+-1)}
\quad{\rm or}\quad
\tilde d_+\xi^{2\tilde q_+-1},
\quad(\xi\to 0), \eqno(37{\rm a})
$$
$$
\tan\eta_-(-E_k)\to d_-\xi^{2q_--1}
\quad{\rm or}\quad
\tilde d_-\xi^{-(2\tilde q_--1)},
\quad(\xi\to 0), \eqno(37{\rm b})
$$
\addtocounter{equation}{1}
where $q_\pm$ and $\tilde q_\pm$ are natural numbers, $d_\pm$ and
$\tilde d_\pm$ are nonzero constants. The first limit in Eq. (37a)
or (37b) corresponds to the case when there exists a critical state
with $E=-\mu$ and corresponding parity. Thus
$\eta_+(-\mu)/\pi$ [$\eta_-(-\mu)/\pi$] takes half integers (integers)
when there exists a critical state with energy $E=-\mu$ and even
(odd) parity, otherwise it takes integers (half integers).

The critical energy states with $E=-\mu$, if exist, are given by
(up to a normalization factor)
\bb
u_{-\mu}^>(x, \pm)=0,
\quad v_{-\mu}^>(x, \pm)=1
\ee    %38
in the region $x>a$. These are also
half-bound states. For a given
potential, the two solutions with $E=-\mu$ cannot apperas
simultaneously.

Using the threshold behaviours obtained above it can be easily
verified that when $k=0$ Eq. (22) reduces to the above critical
solutions or trivial solutions according as the corresponding
critical solutions exist or not.

It may be beneficial to discuss some specific model and to verify
the above threshold behaviour of the phase shifts. A typical and
simple cutoff potential is the square well potential with depth
$V_0$ and width $2a$. This model can be solved explicitly, though
the results are far from simple. Here we are interested only in
scattering states and critical states. The conditions
for the existence of critical energy states can be worked out
explicitly. For scattering states,
one can find the closed form for $\tan\eta_\pm(E_k)$ and
$\tan\eta_\pm(-E_k)$, and discuss the limit $k\to 0$.
Since the calculations
are somewhat tedious but straightforward, we will not give the
details here. We just point out that the general results
and conclusions obtained above
are all confirmed by the simple model at hand.

\section{The Levinson theorem}     %IV

With the above preparations,
we are now ready to establish the Levinson theorem.
The theorem is developed on the basis of the completeness relations
(21) and the threshold behavior of the phase shifts given
in Eqs. (31) and (37). In addition, the fundamental equation (17)
will be employed in the development of the theorem.

Equation (21) is a matrix equation of which the rhs is a diagonal
matrix. When written in matrix elements, it gives four equations.
For free particles, the last two terms on the lhs are absent.
We write down  the first diagonal equation. Replacing $x'$ by $-x'$
in this equation and using Eq. (18) we get another.
Taking the sum and the difference
of these two equations we have
\bb
\int_0^\infty dk\,[u_k^0(x,\pm)u_k^{0*}(x',\pm)+
u_{-k}^0(x,\pm)u_{-k}^{0*}(x',\pm)]=
\frac 12[\delta(x-x')\pm\delta(x+x')],
\ee     %39
where the superscript 0 indicates free particles.
Carrying out the same procedure to the second diagonal 
equation we have
\bb
\int_0^\infty dk\,[v_k^0(x,\pm)v_k^{0*}(x',\pm)+
v_{-k}^0(x,\pm)v_{-k}^{0*}(x',\pm)]=
\frac 12[\delta(x-x')\mp\delta(x+x')].
\ee     %40
The sum of Eqs. (39) and (40) gives
\bb
\int_0^\infty dk\,\sum_s [u_{ks}^0(x,\pm)u_{ks}^{0*}(x',\pm)+
u_{-ks}^0(x,\pm)u_{-ks}^{0*}(x',\pm)]=
\delta(x-x'),
\ee     %41
where the second notation in Eq. (16) has been used and the spinor
index is denoted by $s$. A similar result
for the case  with an external symmetric potential reads
\begin{eqnarray}
&&\int_0^\infty dk\,\sum_s [u_{ks}(x,\pm)u_{ks}^{*}(x',\pm)+
u_{-ks}(x,\pm)u_{-ks}^{*}(x',\pm)] \nonumber\\
&&+\sum_{\kappa_\pm}\sum_s u_{\kappa_\pm s}(x)u_{\kappa_\pm s}^{*}(x')
=\delta(x-x').
\end{eqnarray}     %42
Now we subtract Eq. (42) from Eq. (41), then set $x'=x$, and integrate
over $x$ from $-\infty$ to $+\infty$, we arrive at
\begin{eqnarray}
n_\pm&=&\int_0^\infty dk\,[(\psi_k^0(\pm),\psi_k^0(\pm))-
(\psi_k(\pm),\psi_k(\pm))]   \nonumber\\
&+&\int_0^\infty dk\,[(\psi_{-k}^0(\pm),\psi_{-k}^0(\pm))-
(\psi_{-k}(\pm),\psi_{-k}(\pm))],
\end{eqnarray}    %43
where we have used Eq. (20) to get
$\sum_{\kappa_\pm}(\psi_{\kappa_\pm},\psi_{\kappa_\pm})
=\sum_{\kappa_\pm}1=n_\pm$
where $n_+$ ($n_-$) is the number of bound states with even (odd)
parity. The inner products are defined by integrals similar to
that in Eq. (20). From orthonormal relations we have, say,
$(\psi_k^0(\pm),\psi_k^0(\pm))=\delta(0)\pm\delta(2k)$, and similarly
for the other inner products in Eq. (43). These are infinities and
should be treated very carefully. In order to avoid the difficulty
of infiniteness, we define
\bb
(\psi_{k'}(\pm),\psi_k(\pm))_{r_0}\equiv
\int_{-r_0}^{r_0} dx\,\psi_{k'}^\dagger(x,\pm)\psi_k(x,\pm),
\ee     %44
and obtain $(\psi_k(\pm),\psi_k(\pm))$ in the limit $k'\to k$ and
$r_0\to\infty$. The other inner products in Eq. (43) will be treated
in the same way. Using Eq. (17) it can be shown that
\bb
(\psi_{\pm k'},\psi_{\pm k})_{r_0}=\pm\left.{1\over E_{k'}-E_k}
(v_{\pm k'}^*u_{\pm k}-u_{\pm k'}^*v_{\pm k})\right|_{-r_0}^{r_0},
\ee     %45
which holds for both even-parity and odd-parity solutions. Using the
asymptotic forms (22), and taking the limit $k'\to k$, we find
$$
(\psi_{\pm k}(+),\psi_{\pm k}(+))_{r_0}={r_0\over\pi}+{1\over\pi}
{d\eta_+(\pm E_k)\over dk}\pm
{\mu\over 2\pi kE_k}\sin[2kr_0+2\eta_+(\pm E_k)],
\eqno(46{\rm a})$$
$$
(\psi_{\pm k}(-),\psi_{\pm k}(-))_{r_0}={r_0\over\pi}+{1\over\pi}
{d\eta_-(\pm E_k)\over dk}\mp
{\mu\over 2\pi kE_k}\sin[2kr_0+2\eta_-(\pm E_k)].
\eqno(46{\rm b})$$
For free particles, the corresponding results are obtained by
setting the phase shifts to zero in the above equations.
Obviously, the infiniteness in these results lies in the first term
$r_0/\pi$ in each equation when the limit $r_0\to\infty$ is taken.
This disappears when we take the difference of Eq. (46) and
the corresponding results for free particles.
Using the well-known formulas
$$
\lim_{r_0\to\infty}{\sin 2kr_0\over \pi k}=\delta(k),
$$
and $g(k)\delta(k)=g(0)\delta(k)$ for any continuous function $g(k)$,
we obtain
$$
(\psi_{\pm k}^0(+),\psi_{\pm k}^0(+))_{r_0}-
(\psi_{\pm k}(+),\psi_{\pm k}(+))_{r_0}$$
$$
=-{1\over\pi}{d\eta_+(\pm E_k)\over dk}\pm
\sin^2\eta_+(\pm\mu)\delta(k)\mp
{\mu\over2\pi kE_k}\cos 2kr_0\sin 2\eta_+(\pm E_k),
\eqno(47{\rm a})$$
$$
(\psi_{\pm k}^0(-),\psi_{\pm k}^0(-))_{r_0}-
(\psi_{\pm k}(-),\psi_{\pm k}(-))_{r_0}$$
$$
=-{1\over\pi}{d\eta_-(\pm E_k)\over dk}\mp
\sin^2\eta_-(\pm\mu)\delta(k)\pm
{\mu\over2\pi kE_k}\cos 2kr_0\sin 2\eta_-(\pm E_k).
\eqno(47{\rm b})$$
We regroup the four equations contained in Eq. (47) according to the
sign of the energy instead of the parity, then integrate each group
over $k$ from 0 to $+\infty$, and take the limit $r_0\to\infty$,
we have
$$
\int_0^\infty dk\,[(\psi_{k}^0(\pm), \psi_{k}^0(\pm))-
(\psi_{k}(\pm), \psi_{k}(\pm))] $$
$$
={1\over\pi}[\eta_\pm(\mu)-\eta_\pm(+\infty)]\pm \frac 12
\sin^2\eta_\pm(\mu)\mp
{\mu\over2\pi}\lim_{r_0\to\infty}\int_0^\infty dk\,
{\sin 2\eta_\pm(E_k)\over kE_k}\cos 2kr_0,
\eqno(48{\rm a})$$
$$
\int_0^\infty dk\,[(\psi_{-k}^0(\pm), \psi_{-k}^0(\pm))-
(\psi_{-k}(\pm), \psi_{-k}(\pm))]$$
$$
={1\over\pi}[\eta_\pm(-\mu)-\eta_\pm(-\infty)]\mp \frac 12
\sin^2\eta_\pm(-\mu)\pm
{\mu\over2\pi}\lim_{r_0\to\infty}\int_0^\infty dk\,
{\sin 2\eta_\pm(-E_k)\over kE_k}\cos 2kr_0,
\eqno(48{\rm b})$$
where we have used the integral $\int_0^\infty dk\,\delta(k)=(1/2)
\int_{-\infty}^{+\infty} dk\,\delta(k)=1/2$ since the Dirac $\delta$
function is an even function. So far in this section we have not cut
off the potential. In the following we set $V(x)=V_a(x)$. Then we
have the threshold behavior (31) and (37).
The last term in Eq. (48a) can be decomposed into two integrals,
the first from 0 to $\varepsilon=0^+$ while the second
from $\varepsilon$ to $+\infty$.
The second integral vanishes in the limit $r_0\to\infty$
since the factor $\cos2kr_0$ oscillates very rapidly
and the other factors in the integrand are finite.
For the first integral, we have
\addtocounter{equation}{3}
\bb
\int_0^\varepsilon dk\,
{\sin 2\eta_\pm(E_k)\over kE_k}\cos 2kr_0={1\over \mu}
\int_0^\varepsilon dk\,
{\sin 2\eta_\pm(E_k)\over k}={1\over \mu}
\int_0^{\varepsilon a} d\xi\,{\sin 2\eta_\pm(E_k)\over \xi},
\ee
since $k\le\varepsilon$ is very small. On account of Eq. (31), we have
$\sin 2\eta_\pm(E_k)\to c_\pm\xi^{2r_\pm -1}$ ($\xi\to 0$),
where $r_\pm$ are natural numbers and $c_\pm$ are nonzero constants.
Substituting into the above equation we have
\bb
\int_0^{\varepsilon a} d\xi\,{\sin 2\eta_\pm(E_k)\over \xi}
={c_\pm (\varepsilon a)^{2r_\pm-1}\over 2r_\pm-1}\to 0,
\quad (\varepsilon\to 0^+).
\ee
Thus the first integral vanishes as well. Then the last term in Eq.
(48a) vanishes. By using Eq. (37) it can be shown in a similar way
that the last term in Eq. (48b) also vanish. Substituting the results
(48a, b), each without the last term, into Eq. (43) we arrive at
\begin{equation}
[\eta_\pm(\mu)-\eta_\pm(+\infty)]+[\eta_\pm(-\mu)-\eta_\pm(-\infty)]
\pm {\pi\over2}[\sin^2\eta_\pm(\mu)-\sin^2\eta_\pm(-\mu)]
=n_\pm\pi.
\end{equation}      %51
This is the Levinson theorem for Dirac particles in an external
symmetric potential $V_a(x)$ in one dimension. It relates  the phase
shifts to the total number of bound states for each parity.
In the next section we  discuss some relevant
problems and study two examples.
Finally we summarize the results briefly.

\section{Discussions}           %V
                                    
{\it 1. Phase shifts at infinite momentum}. It should be pointed out
that there is no modulo-$\pi$ ambiguity in Eq. (51), because only
difference of phase shifts at different momentums and trigonometric
functions of the phase shifts are involved. However, it may be
beneficial to appropriately define the phase shifts such that they
can be determined uniquely. We consider the potential $\theta V(x)$
where $0\le\theta\le1$ is a parameter independent of $x$, and $V(x)$
is symmetric but not necessarily be $V_a(x)$. Positive-energy
scattering solutions in this potential will be denoted by
$\psi_k(x,\pm,\theta)$ and negative-energy ones by
$\psi_{-k}(x,\pm,\theta)$. The corresponding phase shifts will be
denoted by $\eta_\pm(E_k,\theta)$ and $\eta_\pm(-E_k,\theta)$
respectively. It is natural to define
\bb
\eta_\pm(E_k,0)=0,\quad \eta_\pm(-E_k,0)=0,
\ee      %52
since there is no potential in this case. We also require that
$\eta_\pm(E_k,\theta)$ and $\eta_\pm(-E_k,\theta)$ be continuous
functions of $\theta$ for any finite $k$. Then 
the phase shifts $\eta_\pm(E_k)=\eta_\pm(E_k,1)$,
$\eta_\pm(-E_k)=\eta_\pm(-E_k,1)$ in the potential $V(x)$ are
definitely defined. It should be remarked that the phase shifts
at threshold are not continuous in $\theta$, however (this was
discussed in some detail in Ref. [18].).

To determine the phase shifts at infinite momentum, we use Eq. (17)
and Eq. (22) for two potentials $\theta V(x)$ and
$\tilde \theta V(x)$. It can be shown that
\bb
\sin[\Delta\eta_+(\pm E_k,\theta)]=\mp\Delta\theta{\pi E_k\over k}
(\psi_{\pm k}(+,\tilde\theta), V\psi_{\pm k}(+,\theta)),
\ee        %53
where $\Delta\theta=\tilde\theta-\theta$ and
$\Delta\eta_+(\pm E_k,\theta)=\eta_+(\pm E_k,\tilde\theta)-
\eta_+(\pm E_k,\theta)$. In the limit $\tilde\theta\to\theta$
or $\Delta\theta\to 0$, the $\psi_{\pm k}(x,+,\tilde\theta)$ in Eq.
(53) can be replaced by $\psi_{\pm k}(x,+,\theta)$, and the sine
on the lhs can be replaced by its argument since the phase shifts
are continuous in $\theta$ as required above, so we have
\begin{equation}
{d\eta_+(\pm E_k,\theta)\over d\theta}=\mp{\pi E_k\over k}
(\psi_{\pm k}(+,\theta),V\psi_{\pm k}(+,\theta)).
\end{equation}    %54
Integrating over $\theta$ from 0 to 1 and using Eq. (52) we have
\bb
\eta_+(\pm E_k)=\mp{\pi E_k\over k} \int_0^1 d\theta\,
(\psi_{\pm k}(+,\theta), V\psi_{\pm k}(+,\theta)).
\ee          %55
For $\eta_-(\pm E_k)$ we have a similar result.
These results are practically not useful since the solutions on the
rhs are not explicitly available in general. However, when $k\to
\infty$, we can ignore the potential $\theta V(x)$ in Eq. (17) since
$V(x)$ is not very singular at $x=0$ (less singular than $x^{-1}$ as
assumed) and is regular elsewhere. Then the solutions $\psi_{\pm k}
(x,+,\theta)$ in Eq. (55) can be replaced by the free ones
$\psi_{\pm k}^0(x,+)$ and we have
\bb
\eta_+(\pm\infty)=\mp\int_0^\infty dx\,V(x).
\ee          %56
The result for
$\eta_-(\pm\infty)$ is the same.
The integral in the above equation converges because
$V(x)$ decreases more rapidly than $x^{-2}$ when $x\to\infty$  and
is less singular than $x^{-1}$ when $x\to 0$, as assumed in Sec. II.
These results are similar to those obtained in three dimensions
[9, 16, 17] and two dimensions [18].
Of course, they hold in the special case $V(x)=V_a(x)$. 
As a consequence, we have
\begin{equation}
\eta_\pm(+\infty)+\eta_\pm(-\infty)=0,
\end{equation}     %57
and the Levinson theorem (51) reduces to the form of Eq. (7).

We have calculated $\eta_+(\pm\infty)$ and $\eta_-(\pm\infty)$
exactly for the square well potential, and the results (56) are
confirmed. In the following we will see that Eq. (56) does not
hold for the $\delta$ potential well, however. This is because
the $\delta$ potential well is more singular than the type we have
assumed. Indeed, if $V(x)$ is less singular than $x^{-1}$ at $x=0$,
or behaves like $A_0x^{-1+\rho/2}$, we have
$\int_0^{\varepsilon} dx\,V(x)=2A_0\varepsilon^{\rho/2}/\rho\to0$
($\varepsilon\to 0^+$). If $V(x)=-U_0\delta(x)$, however, we have
$\int_0^{\varepsilon} dx\,V(x)=-U_0/2$. Thus the
$\delta$ potential well is more singular. It is expected that Eq. (56)
remains correct as long as $V(x)$ belongs to the type we assumed,
i.e., less singular than $x^{-1}$ at $x=0$.

{\it 2. Verification of the theorem.} To examine the Levinson theorem
one should choose a simple potential such that $n_\pm$ and all phase
shifts at zero momentum can be worked out explicitly. This is not
available even for the square well potential.  Although closed forms
for $\tan\eta_\pm(E_k)$ and $\tan\eta_\pm(-E_k)$ can be obtained,
they depend on $k$ in a very complicated way. To determine
$\eta_\pm(\mu)$ and $\eta_\pm(-\mu)$, numerical calculations are
necessary. The transcendental equations for the energy levels of bound
states are also complicated. Perhaps the simplest potential is the
$\delta$ potential well
$V(x)=-U_0\delta(x)$,
where $U_0>0$ is a dimensionless parameter. We have pointed out above
that this potential does not belong to the type we have assumed in
developing the Levinson theorem. However, threshold behavior of the
phase shifts similar to Eqs. (31) and (37) can be explicitly shown.
So the Levinson theorem should remain correct in this case.
As the calculations are simple, we only give the results.
We have
$n_+=1$, $n_-=0$.
The phase shifts at infinite momentum are different from those
given by Eq. (56):
\bb
\eta_+(\pm \infty)=\pm\arctan{U_0\over 2},
\ee     %58
where $\arctan(U_0/2)\in(0,\pi/2)$ is the principal value. A similar
result holds for  $\eta_-(\pm \infty)$. Thus Eq. (57) remains valid
in this case though Eq. (56) does not, and the reduced form of the
Levinson theorem (7) is expected to hold. In fact,
the phase shifts at threshold can be found to be
\bb
\eta_+(\mu)={\pi\over 2},\quad \eta_+(-\mu)=0,\quad
\eta_-(\mu)=0,\quad \eta_-(-\mu)=-{\pi\over 2},
\ee     %59
and it is easy to verify that Eq. (7) is
satisfied.

For a $\delta$ potential barrier $V(x)=U_0\delta(x)$ where
$U_0>0$, it can be shown that $n_+=0$, $n_-=1$,
$\eta_+(\mu)=-\pi/2$, $\eta_+(-\mu)=0$,
$\eta_-(\mu)=0$,  $\eta_-(-\mu)=\pi/2$, and Eq. (7) is satisfied
as well.

A less simple example is the double $\delta$ potential wells
$V(x)=-U_0[\delta(x-a)+\delta(x+a)]$,
where $U_0>0$ is dimensionless. This example is somewhat more
substantial since half-bound states may be involved,
and both $n_\pm$ and the phase shifts depend on the value of $U_0$.
Since the calculations are not difficult but lengthy, we will not
give the details. We only point out that in this case the
Levinson theorem is confirmed once again.

{\it 3. Summary}. In this paper we study Dirac particles in one-%
dimensional symmetric potentials that decrease more rapidly than
$x^{-2}$ when $x\to\infty$ and are less singular than $x^{-1}$ when
$x\to 0$. The properties of bound state and scattering solutions 
are discussed. For cutoff potentials the threshold behaviours of the
phase shifts are studied in detail, and the Levinson theorem (7) is
established, which connects the phase shifts with the total number of
bound states for each parity. Two simple examples are discussed and
the Levinson theorem is verified explicitly. A mathematically
rigorous extension of the Levinson theorem to non-cutoff potentials
is still not available. However, one may expect that the theorem
remains correct for short-range non-cutoff potentials where the
asymptotic forms (24) for scattering solutions hold and the total
number of bound states is finite.

\section*{Acknowledgment}

The author is grateful to Professor Guang-jiong Ni for 
discussions and encouragement.
This work was supported by the
National Natural Science Foundation of China.

%\newpage

\end{document}